\documentclass{aastex631}

\newcommand\years{7.5 }

\usepackage{textcomp}
\graphicspath{{./}{figures/}}

\usepackage{multirow}
\usepackage{lipsum}
\usepackage{threeparttable}
\usepackage{comment} 

\usepackage{soul}

\usepackage{silence}
\shorttitle{Radio Galaxies with HAWC}
\usepackage{CJKutf8}

\begin{document}

\title{Longtime Monitoring of TeV Radio Galaxies with HAWC}

\correspondingauthor{D.~Avila Rojas }
\email{daniel\_avila5@ciencias.unam.mx}
\correspondingauthor{T.~Capistrán}
\email{tomas.capistranrojas@unito.it}
\correspondingauthor{M.M.~González}
\email{tcapistran@astro.unam.mx}

\author[0000-0002-4713-7069]{R.~Alfaro}
\affiliation{Instituto de F\'{i}sica, Universidad Nacional Autónoma de México, Ciudad de Mexico, Mexico }

\author{C.~Alvarez}
\affiliation{Universidad Autónoma de Chiapas, Tuxtla Gutiérrez, Chiapas, México}

\author{E.~Anita-Rangel}
\affiliation{Instituto de Astronom\'{i}a, Universidad Nacional Autónoma de México, Ciudad de Mexico, Mexico }

\author{J.C.~Arteaga-Velázquez}
\affiliation{Universidad Michoacana de San Nicolás de Hidalgo, Morelia, Mexico }

\author[0000-0002-4020-4142]{D.~Avila Rojas}
\affiliation{Instituto de Astronom\'{i}a, Universidad Nacional Autónoma de México, Ciudad de Mexico, Mexico }

\author[0000-0002-2084-5049]{H.A.~Ayala Solares}
\affiliation{Department of Physics, Pennsylvania State University, University Park, PA, USA }

\author[0000-0002-5529-6780]{R.~Babu}
\affiliation{Department of Physics and Astronomy, Michigan State University, East Lansing, MI, USA }

\author{P.~Bangale}
\affiliation{Temple University, Department of Physics, 1925 N. 12th Street, Philadelphia, PA 19122, USA}

\author{E.~Belmont-Moreno}
\affiliation{Instituto de F\'{i}sica, Universidad Nacional Autónoma de México, Ciudad de Mexico, Mexico }

\author{A.~Bernal}
\affiliation{Instituto de Astronom\'{i}a, Universidad Nacional Autónoma de México, Ciudad de Mexico, Mexico }

\author[0000-0002-4042-3855]{K.S.~Caballero-Mora}
\affiliation{Universidad Autónoma de Chiapas, Tuxtla Gutiérrez, Chiapas, México}

\author[0000-0003-2158-2292]{T.~Capistrán}
\affiliation{Universit`a degli Studi di Torino, I-10125 Torino, Italy}
\affiliation{Instituto de Astronom\'{i}a, Universidad Nacional Autónoma de México, Ciudad de Mexico, Mexico }

\author[0000-0002-8553-3302]{A.~Carramiñana}
\affiliation{Instituto Nacional de Astrof\'{i}sica, Óptica y Electrónica, Puebla, Mexico }

\author{F.~Carreón}
\affiliation{Instituto de Astronom\'{i}a, Universidad Nacional Autónoma de México, Ciudad de Mexico, Mexico }

\author[0000-0002-6144-9122]{S.~Casanova}
\affiliation{Institute of Nuclear Physics Polish Academy of Sciences, PL-31342 IFJ-PAN, Krakow, Poland }

\author[0000-0002-7607-9582]{U.~Cotti}
\affiliation{Universidad Michoacana de San Nicolás de Hidalgo, Morelia, Mexico }

\author[0000-0002-1132-871X]{J.~Cotzomi}
\affiliation{Facultad de Ciencias F\'{i}sico Matemáticas, Benemérita Universidad Autónoma de Puebla, Puebla, Mexico }

\author[0000-0002-7747-754X]{S.~Coutiño de León}
\affiliation{Dept. of Physics and Wisconsin IceCube Particle Astrophysics Center, University of Wisconsin{\textemdash}Madison, Madison, WI, USA}

\author[0000-0001-9643-4134]{E.~De la Fuente}
\affiliation{Departamento de F\'{i}sica, Centro Universitario de Ciencias Exactase Ingenierias, Universidad de Guadalajara, Guadalajara, Mexico }

\author{D.~Depaoli}
\affiliation{Max-Planck Institute for Nuclear Physics, 69117 Heidelberg, Germany}

\author{P.~Desiati}
\affiliation{Dept. of Physics and Wisconsin IceCube Particle Astrophysics Center, University of Wisconsin{\textemdash}Madison, Madison, WI, USA}

\author{N.~Di Lalla}
\affiliation{Department of Physics, Stanford University: Stanford, CA 94305–4060, USA}

\author{R.~Diaz Hernandez}
\affiliation{Instituto Nacional de Astrof\'{i}sica, Óptica y Electrónica, Puebla, Mexico }

\author[0000-0002-2987-9691]{M.A.~DuVernois}
\affiliation{Dept. of Physics and Wisconsin IceCube Particle Astrophysics Center, University of Wisconsin{\textemdash}Madison, Madison, WI, USA}

\author[0000-0002-0087-0693]{J.C.~Díaz-Vélez}
\affiliation{Dept. of Physics and Wisconsin IceCube Particle Astrophysics Center, University of Wisconsin{\textemdash}Madison, Madison, WI, USA}

\author{K.~Engel}
\affiliation{Department of Physics, University of Maryland, College Park, MD, USA }

\author{T.~Ergin}
\affiliation{Department of Physics and Astronomy, Michigan State University, East Lansing, MI, USA }

\author[0000-0001-7074-1726]{C.~Espinoza}
\affiliation{Instituto de F\'{i}sica, Universidad Nacional Autónoma de México, Ciudad de Mexico, Mexico }

\author[0000-0002-0173-6453]{N.~Fraija}
\affiliation{Instituto de Astronom\'{i}a, Universidad Nacional Autónoma de México, Ciudad de Mexico, Mexico }

\author{S.~Fraija}
\affiliation{Instituto de Astronom\'{i}a, Universidad Nacional Autónoma de México, Ciudad de Mexico, Mexico }

\author[0000-0002-4188-5584]{J.A.~García-González}
\affiliation{Tecnologico de Monterrey, Escuela de Ingenier\'{i}a y Ciencias, Ave. Eugenio Garza Sada 2501, Monterrey, N.L., Mexico, 64849}

\author{F.~Garfias}
\affiliation{Instituto de Astronom\'{i}a, Universidad Nacional Autónoma de México, Ciudad de Mexico, Mexico }

\author{A.~Gonzalez Muñoz}
\affiliation{Instituto de F\'{i}sica, Universidad Nacional Autónoma de México, Ciudad de Mexico, Mexico }

\author[0000-0002-5209-5641]{M.M.~González}
\affiliation{Instituto de Astronom\'{i}a, Universidad Nacional Autónoma de México, Ciudad de Mexico, Mexico }

\author{J.A.~González}
\affiliation{Universidad Michoacana de San Nicolás de Hidalgo, Morelia, Mexico }

\author[0000-0002-9790-1299]{J.A.~Goodman}
\affiliation{Department of Physics, University of Maryland, College Park, MD, USA }

\author{S.~Groetsch}
\affiliation{Department of Physics, Michigan Technological University, Houghton, MI, USA }

\author[0000-0001-9844-2648]{J.P.~Harding}
\affiliation{Los Alamos National Laboratory, Los Alamos, NM, USA }

\author{I.~Herzog}
\affiliation{Department of Physics and Astronomy, Michigan State University, East Lansing, MI, USA }

\author[0000-0002-3808-4639]{D.~Huang}
\affiliation{Department of Physics, University of Maryland, College Park, MD, USA }

\author[0000-0002-5527-7141]{F.~Hueyotl-Zahuantitla}
\affiliation{Universidad Autónoma de Chiapas, Tuxtla Gutiérrez, Chiapas, México}

\author{A.~Iriarte}
\affiliation{Instituto de Astronom\'{i}a, Universidad Nacional Autónoma de México, Ciudad de Mexico, Mexico }

\author{S.~Kaufmann}
\affiliation{Universidad Politecnica de Pachuca, Pachuca, Hgo, Mexico }

\author{D.~Kieda}
\affiliation{Department of Physics and Astronomy, University of Utah, Salt Lake City, UT, USA }

\author{A.~Lara}
\affiliation{Instituto de Geof\'{i}sica, Universidad Nacional Autónoma de México, Ciudad de Mexico, Mexico }

\author{J.~Lee}
\affiliation{University of Seoul, Seoul, Rep. of Korea}

\author[0000-0001-5516-4975]{H.~León Vargas}
\affiliation{Instituto de F\'{i}sica, Universidad Nacional Autónoma de México, Ciudad de Mexico, Mexico }

\author{[0000-0001-8825-3624]A.L.~Longinotti}
\affiliation{Instituto de Astronom\'{i}a, Universidad Nacional Autónoma de México, Ciudad de Mexico, Mexico }

\author[0000-0003-2810-4867]{G.~Luis-Raya}
\affiliation{Universidad Politecnica de Pachuca, Pachuca, Hgo, Mexico }

\author[0000-0001-8088-400X]{K.~Malone}
\affiliation{Los Alamos National Laboratory, Los Alamos, NM, USA }

\author[0000-0001-9052-856X]{O.~Martinez}
\affiliation{Facultad de Ciencias F\'{i}sico Matemáticas, Benemérita Universidad Autónoma de Puebla, Puebla, Mexico }

\author[0000-0002-2824-3544]{J.~Martínez-Castro}
\affiliation{Centro de Investigaci\'on en Computaci\'on, Instituto Polit\'ecnico Nacional, M\'exico City, M\'exico.}

\author[0000-0002-2610-863X]{J.A.~Matthews}
\affiliation{Dept of Physics and Astronomy, University of New Mexico, Albuquerque, NM, USA }

\author[0000-0002-8390-9011]{P.~Miranda-Romagnoli}
\affiliation{Universidad Autónoma del Estado de Hidalgo, Pachuca, Mexico }

\author{J.A.~Morales-Soto}
\affiliation{Universidad Michoacana de San Nicolás de Hidalgo, Morelia, Mexico }

\author[0000-0002-1114-2640]{E.~Moreno}
\affiliation{Facultad de Ciencias F\'{i}sico Matemáticas, Benemérita Universidad Autónoma de Puebla, Puebla, Mexico }

\author[0000-0002-7675-4656]{M.~Mostafá}
\affiliation{Temple University, Department of Physics, 1925 N. 12th Street, Philadelphia, PA 19122, USA}

\author{M.~Najafi}
\affiliation{Department of Physics, Michigan Technological University, Houghton, MI, USA }

\author{A.~Nayerhoda}
\affiliation{Institute of Nuclear Physics Polish Academy of Sciences, PL-31342 IFJ-PAN, Krakow, Poland }

\author[0000-0003-1059-8731]{L.~Nellen}
\affiliation{Instituto de Ciencias Nucleares, Universidad Nacional Autónoma de Mexico, Ciudad de Mexico, Mexico }

\author{N.~Omodei}
\affiliation{Department of Physics, Stanford University: Stanford, CA 94305–4060, USA}

\author[0009-0009-2481-6921]{M.~Osorio-Archila}
\affiliation{Instituto de Astronom\'{i}a, Universidad Nacional Autónoma de México, Ciudad de Mexico, Mexico }

\author{E.~Ponce}
\affiliation{Facultad de Ciencias F\'{i}sico Matemáticas, Benemérita Universidad Autónoma de Puebla, Puebla, Mexico }

\author{Y.~Pérez Araujo}
\affiliation{Instituto de F\'{i}sica, Universidad Nacional Autónoma de México, Ciudad de Mexico, Mexico }

\author{E.G.~Pérez-Pérez}
\affiliation{Universidad Politecnica de Pachuca, Pachuca, Hgo, Mexico }

\author[0000-0002-6524-9769]{C.D.~Rho}
\affiliation{Department of Physics, Sungkyunkwan University, Suwon 16419, South Korea}

\author{A.~Rodriguez Parra}
\affiliation{Universidad Michoacana de San Nicolás de Hidalgo, Morelia, Mexico }

\author[0000-0003-1327-0838]{D.~Rosa-González}
\affiliation{Instituto Nacional de Astrof\'{i}sica, Óptica y Electrónica, Puebla, Mexico }

\author{M.~Roth}
\affiliation{Los Alamos National Laboratory, Los Alamos, NM, USA }

\author{H.~Salazar}
\affiliation{Facultad de Ciencias F\'{i}sico Matemáticas, Benemérita Universidad Autónoma de Puebla, Puebla, Mexico }

\author[0000-0001-6079-2722]{A.~Sandoval}
\affiliation{Instituto de F\'{i}sica, Universidad Nacional Autónoma de México, Ciudad de Mexico, Mexico }

\author[0000-0001-8644-4734]{M.~Schneider}
\affiliation{Department of Physics, University of Maryland, College Park, MD, USA }

\author[0009-0007-7409-4061]{J.~Serna-Franco}
\affiliation{Instituto de F\'{i}sica, Universidad Nacional Autónoma de México, Ciudad de Mexico, Mexico }

\author{Y.~Son}
\affiliation{University of Seoul, Seoul, Rep. of Korea}

\author[0000-0002-1492-0380]{R.W.~Springer}
\affiliation{Department of Physics and Astronomy, University of Utah, Salt Lake City, UT, USA }

\author{O.~Tibolla}
\affiliation{Universidad Politecnica de Pachuca, Pachuca, Hgo, Mexico }

\author[0000-0001-9725-1479]{K.~Tollefson}
\affiliation{Department of Physics and Astronomy, Michigan State University, East Lansing, MI, USA }

\author[0000-0002-1689-3945]{I.~Torres}
\affiliation{Instituto Nacional de Astrof\'{i}sica, Óptica y Electrónica, Puebla, Mexico }

\author{R.~Torres-Escobedo}
\affiliation{Tsung-Dao Lee Institute \& School of Physics and Astronomy, Shanghai Jiao Tong University, 800 Dongchuan Rd, Shanghai, SH 200240, China}

\author{E.~Varela}
\affiliation{Facultad de Ciencias F\'{i}sico Matemáticas, Benemérita Universidad Autónoma de Puebla, Puebla, Mexico }

\author{L.~Villaseñor}
\affiliation{Facultad de Ciencias F\'{i}sico Matemáticas, Benemérita Universidad Autónoma de Puebla, Puebla, Mexico }

\author{X.~Wang}
\affiliation{Department of Physics, Michigan Technological University, Houghton, MI, USA }

\author{Z.~Wang}
\affiliation{Department of Physics, University of Maryland, College Park, MD, USA }

\author{I.J.~Watson}
\affiliation{University of Seoul, Seoul, Rep. of Korea}

\author{H.~Wu}
\affiliation{Dept. of Physics and Wisconsin IceCube Particle Astrophysics Center, University of Wisconsin{\textemdash}Madison, Madison, WI, USA}

\author{S.~Yu}
\affiliation{Department of Physics, Pennsylvania State University, University Park, PA, USA }

\author[0000-0003-0513-3841]{H.~Zhou}
\affiliation{Tsung-Dao Lee Institute \& School of Physics and Astronomy, Shanghai Jiao Tong University, 800 Dongchuan Rd, Shanghai, SH 200240, China}

\author[0000-0002-8528-9573]{C.~de León}
\affiliation{Universidad Michoacana de San Nicolás de Hidalgo, Morelia, Mexico }





\begin{abstract}

We present the monitoring of the TeV-emitting radio galaxies M87, NGC~1275, 3C~264, and IC~310 with the High Altitude Water Cherenkov Observatory (HAWC) over a period of approximately $7.5$ years. The analysis includes light curves at daily, weekly and monthly time scales for the four sources. We report the detection of gamma-ray emission from M87 with a significance exceeding 5$\sigma$. Due to its significant detection, this work reports the integrated TeV spectrum of M87 from the longest temporal coverage up to date. The source is well described as a point-like source modeled by a power law spectrum with spectral index $\alpha = 2.53\pm0.29$ and a flux of  $(7.09\pm 1.24)\times10^{-13}$ $\rm{cm}^{-2}\,{s}^{-1}\,{TeV}^{-1}$ at $1\,\rm{TeV}$. The maximum energy of the detected emission in M87, at 1$\sigma$ confidence level (C.L.), reaches 26.5 TeV. HAWC's observation of M87 reveals a low flux spectrum for the longest observation to date of this radio galaxy. 3C~264 is marginally detected with a significance slightly below 4$\sigma$, while NGC~1275 and IC~310 are not detected. The weekly light curves show an increased number of fluxes above $2\sigma$ for M87 starting in 2019, and for 3C~264 starting in 2018, which can be interpreted as the moment for which these sources start to exhibit an enhanced steady TeV emission. Overall, in the four radio galaxies, the cumulative significance over time indicates a behavior that resembles that of a gamma-ray variable active galaxy, such as the blazar Markarian 421. This supports the importance of monitoring radio galaxies to identify periods of higher activity and flares, enabling further multi-messenger studies.

\end{abstract}



\section{Introduction}\label{sec:intro}

Radio galaxies represent a sub-type of radio-loud Active Galactic Nuclei (AGNs) with their jets misaligned from our line of sight. They emit non-thermal radiation across a broad spectrum, from radio to gamma-rays, originating in their core, jet, and lobes. These sources offer a unique perspective on the acceleration mechanisms operating within their jets due to their misaligned geometry and the lack of of strong Doppler-boosted emission, distinguishing them from blazars whose jet structures are unobservable because they are oriented close to our line-of-sight. The detection of radio galaxies at TeV energies has prompted significant investigations into the dynamics and acceleration mechanisms of AGN jets \citep{2006Sci...314.1424A,2012A&A...539L...2A,aleksic2014rapid,2020ApJ...896...41A}. This represents an opportunity to explore the processes and locations of gamma-ray emissions, the presence of multiple gamma-ray emission components, and whether they originate near the black hole or along the jet ($0.1-1$ pc from the black hole) \citep{2022Galax..10...61R}. Several key findings have emerged, including the influence of a clumpy environment on gamma-ray emission and the occurrence of orphan flares \citep{2018ApJ...864..118K}. The study of different emission components within radio galaxies and their association with gamma-ray emission is intriguing, leading to numerous efforts to determine the most suitable spectral models to describe the emission from these sources. 
    
Their Spectral Energy Distribution (SED) can be described with leptonic, hadronic or lepto-hadronic models. These models can explain emissions up to TeV energies: leptonic models explain the emission through Synchrotron Self-Compton mechanism (SSC) \citep{1993ApJ...416..458D, 2014A&A...562A..12P}, while hadronic models involve proton-proton or proton-photon interactions \citep{2012PhRvD..85d3012S,2014MNRAS.441.1209F}. Additionally, hadronic models have been used to investigate the relationship between gamma-ray emission and the estimated neutrino flux in radio galaxies \citep{2012JPhCS.378a2005A,2016ApJ...830...81F,2016ApJ...833..279B,2022ApJ...934..158A}. It has been proposed that the neutrino flux produced in AGNs, particularly in Fanaroff-Riley I (FR-I) radio galaxies in which the peak of the radio emission is located near the core, may contribute to the diffuse flux of neutrinos in the universe \citep{2015IAUS..313..169M,2012APh....37...40A}. However, it has been suggested that the neutrino flux may be below the current sensitivity thresholds of observatories, resulting in non-detection \citep{2016ApJ...833..279B,2012JPhCS.378a2005A}. In addition, under some unification models, FR-I radio galaxies are considered to be BL Lacs objects with their jet misaligned from our line of sight \citep{1995PASP..107..803U,2012MNRAS.420.2899G,2024Ap&SS.369..106I}.

To date, four FR-I radio galaxies have been detected at TeV energies: M87, Centaurus A (Cen A), NGC~1275, and 3C~264. These radio galaxies have been studied by multiple gamma-ray observatories, investigating  their emission up to a few TeV \citep{de2019hess,aleksic2016major,aleksic2016major2,holder2011veritas,2011ICRC...12..137H}. In addition, IC~310 is a TeV source that was formerly classified as a head-tail radio galaxy \citep{1998A&A...331..901S}, its true nature is still under debate since its nucleus has a blazar-like behavior \citep{2012A&A...538L...1K}. However, high-resolution observations with the VLA at low frequencies reveal two distinct narrowly collimated jets that are strongly bent by the ram pressure of the inter-cluster medium (ICM), favoring the head-tail radio galaxy scenario for IC~310 \citep{2020MNRAS.499.5791G}. Thus, we consider IC~310 as a TeV FR-I radio galaxy for the present work. A more comprehensive understanding of these sources can be achieved through synergy between different observatories at energies of tens to hundreds of TeV. In particular, a Wide Field-of-View Detector (WFD), such as the HAWC observatory, is able to perform long-term monitoring of sources within its Field-of-View (FoV). 
    
Notably, HAWC's FoV includes four of the previously TeV-detected radio galaxies: M87, NGC~1275, IC~310 and 3C~264. In a previous HAWC analysis, these radio galaxies were studied as part of an AGN survey, with M87 being marginally detected with a TS of 12.93 and average emission consistent with an intermediate activity state; while NGC~1275, IC~310 and 3C~264 were not significantly detected \citep{2021ApJ...907...67A}. The forthcoming work will examine the most recent results obtained with the HAWC observatory after \years years of observations of the aforementioned four radio galaxies using the advanced capabilities of the HAWC observatory and employing the latest event reconstruction algorithms, known as PASS~5 \citep{2024ApJ...972..144A}, which enhanced the reconstruction and background rejection at lower energies and the sensitivity and angular resolution at higher energies. This represents the most contemporary and extensive monitoring of radio galaxies at TeV energies. A brief summary of the past TeV observations for each of the radio galaxies will be presented.

\paragraph*{}{\textbf{M87}}

    M87 is a giant elliptical radio galaxy in the Virgo cluster, located at $\alpha =187.65^{\circ}$, $ \delta = 12.396^{\circ} $ in equatorial coordinates (EqJ2000.0). At its core lies a supermassive black hole (SMBH) with a mass of $ 6.5 \pm 0.7 \times 10^{9} \; \rm{M}_{\odot} $ \citep{2019ApJ...875L...1E} and exhibits a relativistic jet extending from $ 1.5$ to $ 2 \; \rm{kpc} $ with an estimated angle of $ 15^{\circ} - 25^{\circ} $ relative to the line of sight \citep{2009Sci...325..444A}. At a redshift of $z = 0.0044$ and a redshift-independent measured distance from Earth of $ 16.7 \pm 0.2 \; \rm{Mpc} $ \citep{2007ApJ...655..144M}, it is the second closest radio galaxy known, but the closest within HAWC's FoV. Due to its proximity and the misalignment of its jet, extensive high-resolution studies of the jet structure have been conducted across various frequency bands, ranging from radio to X-rays \citep{2009Sci...325..444A,2023MNRAS.526.5949N,2021ApJ...919..110I}. Furthermore, in the radio band, the Event Horizon Telescope (EHT) achieved a groundbreaking milestone by capturing the first-ever image of a SMBH in M87's core \citep{2019ApJ...875L...1E,2019ApJ...875L...2E,2019ApJ...875L...3E,2019ApJ...875L...4E,2019ApJ...875L...5E,2019ApJ...875L...6E}. Regarding the very-high energy (VHE) gamma-ray band, M87 provides a unique opportunity to study its emission, characterize its SED, and explore its particle acceleration processes, with the aim of generalizing the findings to radio galaxies and AGNs \citep{2022ApJ...938...79B, 2023A&A...675A.138H}.
    
    The first detection of M87 above 730 $\rm{GeV}$ was made by HEGRA, which reported an equivalent flux of $(3.3 \pm 0.8) \%$ of the Crab Nebula flux \citep{2003A&A...403L...1A}. Subsequently, numerous Imaging Atmospheric Cherenkov Telescopes (IACTs) have monitored M87, registering both quiescent and active states during their observational campaigns. In quiescent state, M87 has been observed by H.E.S.S., MAGIC and VERITAS \citep{2006Sci...314.1424A,2012A&A...544A..96A,2008ApJ...679..397A}. The 2004 H.E.S.S. observation recorded the lowest flux ever measured for this source \citep{2006Sci...314.1424A}. The spectrum observed for this state is more likely to follow a Simple Power Law (SPL) (see row 1 in Table \ref{tab:RadGals}). During the 2011-2012 campaign, VERITAS observed monthly variations in the TeV flux, suggesting that the quiescent state evolves over longer time scales compared to flare states \citep{2012AIPC.1505..586B}. The spectral analysis was derived using data from two months of elevated flux \citep{2012AIPC.1505..586B} (see row 4 and 5 in Table \ref{tab:RadGals}). For high-activity states, flares have been observed in 2005, 2008, and 2010 by H.E.S.S., MAGIC and VERITAS, respectively \citep{2006Sci...314.1424A, 2008ApJ...685L..23A, 2010ApJ...716..819A,2012ApJ...746..141A}. The spectrum in these high-activity states were fitted to a SPL in all observations. 
    
    MAGIC and VERITAS conducted a monitoring of M87 covering observations from 2019 to 2022, with a total of 112 and 61 hours of effective observation time, respectively. Both, daily and monthly scale light curves did not exhibit variability, and a joint spectral analysis with Fermi-LAT resulted in a power law with index of $2.28 \pm 0.02$ \citep{MoleroGonzalez:2023aez}. On the other hand, H.E.S.S. analyzed data from 2004 to 2021, accumulating a total of 120 hours of data during low states of the source. H.E.S.S analysis revealed that the morphology of the gamma-ray emissions in the different activity states is compatible with the radio core \citep{Arcaro:20232B}. The spectral fitted parameters for some of the main observations of this radio galaxy at different activity states are summarized in Table \ref{tab:RadGals}. These findings underscore the importance of M87 as a milestone target in the study of TeV-emitting radio galaxies and AGNs. 
    
    Regarding WFD, LHAASO performed observations of M87 from 2021 to 2024, detecting VHE gamma-ray emission with a statistical significance of approximately $9\sigma$. An 8-day VHE flare was found in LHAASO's data in early 2022, with a rise time of $\sim1.05$ days and a decay time of ~2.17 days, suggesting a compact emission region of only a few Schwarzschild radii of the central supermassive black hole \citep{2024ApJ...975L..44C}. On the other hand, the HAWC observatory on previous investigations using different datasets did not formally detected M87, only marginally \citep{2021ApJ...907...67A}, but conducted a comprehensive analysis of its SED at higher energy ranges using HAWC upperlimits, where the long-term TeV emission of M87 was interpreted within the framework of a lepto-hadronic model capable to also explain M87's 2005 orphan flare episode. \citep{2022ApJ...934..158A}.
    
\paragraph*{}{\bf NGC~1275}

    NGC~1275, the most luminous galaxy in the Perseus cluster, is situated at a distance of approximately 75 Mpc, as measured independently of its redshift $z = 0.0175$ \citep{2022Galax..10...61R}, and located at ($\alpha,\delta$) = ($49.95^\circ,41.51^\circ$) in equatorial coordinates \citep{2014ApJS..212...18B}. At its core resides a SMBH with an estimated mass of $3-4 \times 10^{8} \; \rm{M}_{\odot}$ \citep{2005MNRAS.359..755W}. Numerous radio observations have revealed a sub-parsec-scale radio jet, which is oriented between approximately $30^\circ$ and $60^\circ$ with respect to the line of sight \citep{2018ApJ...864..118K,2021ApJ...920L..24K}. TeV emission from NGC~1275 has been detected by MAGIC \citep{2012A&A...539L...2A} and VERITAS \citep{2019ICRC...36..632B}. MAGIC made the initial discovery above 100 GeV with a confidence level of $6\sigma$ during two observation campaigns conducted from October 2009 to February 2010 and from August 2010 to February 2011, resulting in a total data collection of approximately 100 hours. In the last campaign, the MAGIC Collaboration characterized NGC~1275's spectrum with a SPL; the corresponding spectral index and flux normalization values are shown on row 11 in Table \ref{tab:RadGals}.
    
    Two flares in high-energy (HE) gamma rays ($>100$ MeV) were reported by Fermi-LAT in October 2015 and December 2016/January 2017 \citep{2017ApJ...848..111B}. In both instances, variability was detected on hour scales, indicating a highly compact emission region. Furthermore, two periods of VHE activity have been reported in NGC~1275 since 2016. The first occurred at the end of October 2016, during which the flux increased up to 16\% of the Crab Nebula flux \citep{2016ATel.9689....1M}. The second occurred on 31 December 2016, reaching a flux approximately 1.5 times that of the Crab Nebula flux, representing the highest flare observed from this source \citep{2017ATel.9929....1M, 2018A&A...617A..91M}. For the 31 December 2016 flare, the MAGIC Collaboration reports a power law with exponential cutoff (PLEC) for the fitted spectra. No flux was observed beyond $ 650 \; \rm{GeV}$ suggesting a cutoff in its spectrum (see row 13 in Table \ref{tab:RadGals}). In this same analysis, a study was carried out including the Fermi-LAT data at very high energies, with the best fit to the spectrum being a PLEC \citep{2018A&A...617A..91M}. 

\paragraph*{}{\bf IC~310}

    IC~310 is an AGN also located in the Perseus cluster near NGC~1275 at $(\alpha,\delta) = (49.179^\circ,41.325^\circ)$ with a redshift of $z = 0.0189$ \citep{2002AJ....123.2990B}, at a distance of $\sim 80\,\rm{Mpc}$ from Earth \citep{2010A&A...519L...6N}. At TeV energies, IC~310 has been detected in different activity states by MAGIC \citep{aleksic2014rapid,2017A&A...603A..25A} with a notably hard spectra (spectral index $\sim-2$), independently of the level of activity. MAGIC reported flare activity occurring in November 2012 and the data is well-fitted by a point-like source with a power law spectrum. The most recent flare reported from this source was detected on March 2024 by LHAASO, with a flux reaching approximately 0.5 Crab Units above 1 TeV\citep{2024ATel16513....1X}. The values of the best-fit parameters of the spectrum for different MAGIC observational campaigns are reported in Table \ref{tab:RadGals}.
    
\paragraph*{}{\bf 3C~264}

    3C~264 is a radio galaxy within the Leo cluster, estimated to be at a distance of $\sim 95$ Mpc with a redshift of z = 0.022 \citep{1999ApJS..125...35S,2010ApJ...708..171P}, and located at $(\alpha,\delta) = (176.271^\circ, 19.606^\circ)$ in equatorial coordinates \citep{2004AJ....127.3587F}. At its core lies a SMBH with an approximate mass of $5\times 10^8 \rm{M}_\odot$ \citep{2015A&A...581A..33D}, and exhibits a relativistic jet extending up to kpc scales, where knots have been observed in the optical band \citep{2015Natur.521..495M}. TeV emission from 3C~264 was first detected by VERITAS in 2018 during a high-activity state. Observations were conducted for approximately 12 hours between February and March 2018. The initial findings indicated an excess of gamma-ray events compared to the background, with a significance of $5.4\sigma$ and an estimated integrated flux of approximately 1\% of the Crab Nebula flux \citep{2018ATel11436....1M}. VERITAS later extended the analysis of the source using data from 2017 to 2019, resulting in a detection with a significance of $7.8\sigma$ \citep{2020ApJ...896...41A}. They reported weak variability on the time scales ranging from months to years \citep{2018Galax...6..116R}, suggesting that the detection occurred in early 2018 and it did not arise from flare activity in the source but rather from a moderately enhanced flux state. The parameters of the spectral model at VHE observed by VERITAS are reported in row 14 in Table \ref{tab:RadGals}. 
    
\begin{table}[ht!]
    \centering
    \caption{Spectral parameter values considering a simple power law (SPL) and a power law with exponential cutoff (PLEC) reported for the radio galaxies M87, NGC~1275, 3C~264 and IC~310 on different activity periods. The Experiment column shows the instrument with which the observation was made, following the year of the observation.}
    \resizebox{\textwidth}{!}{
    \begin{tabular}{cclcccccc}
        \hline
        & \multirow{2}{*}{\textbf{Source}} & \multirow{2}{*}{\textbf{Experiment}} & Activity & \textbf{Fit} & \multirow{2}{*}{\textbf{Spectral Index}} & \textbf{Flux at} $\mathbf{1\;\mathbf{TeV}}$ & $\mathbf{E_{cut}}$  & \multirow{2}{*}{\textbf{Ref}} \\
        & & & State & \textbf{Model} & & [$\mathbf{cm}^{-2}\,\mathbf{s}^{-1}\,\mathbf{TeV}^{-1}$] $\times 10^{-13}$ & [\textbf{TeV}] & \\
        \hline
        1 & \multirow{10}{*}{M87} & H.E.S.S., 2004 & quiescent & SPL & $2.62\pm 0.35$ & $2.43\pm 0.75$ & - & [1] \\ 
        2 & & MAGIC, 2005-2007 & quiescent & SPL & $2.21\pm 0.21$ & $5.4\pm 1.1$ & - & [2] \\
        3 & & VERITAS, 2007 & quiescent & SPL & $2.31 \pm 0.17$ & $7.4\pm 1.3$ & - & [3] \\
        4 & & VERITAS, 2012-1 & quiescent & SPL & $2.1\pm 0.3$ & $6.3\pm 1.6$ & - & [4] \\
        5 & & VERITAS, 2012-2 & quiescent & SPL & $2.6\pm 0.2$ & $7.0\pm 1.5$ & - & [4] \\
        6 & & MAGIC, 2012-2015 & quiescent & SPL & $2.41\pm 0.07$ & $3.95\pm 0.33$ & - & [5] \\ 
        7 & & H.E.S.S., 2005 & flare & SPL & $2.22\pm 0.15$ & $11.7\pm 1.6$ & - & [1] \\
        8 & & MAGIC, 2008 & flare & SPL & $2.21\pm 0.18$ & $48.1\pm 8.2$ & - & [6] \\ 
        9 & & VERITAS, 2008-2009 & flare & SPL & $2.40\pm 0.21$ & $15.9\pm 2.9$ & - & [7] \\ 
        10 & & VERITAS, 2010 & flare & SPL & $2.19\pm 0.07$ & $47.1\pm 2.9$ & - & [8] \\
        \hline 
        11 & \multirow{3}{*}{NGC~1275} & MAGIC, 2010-2011$^*$ & quiescent & SPL & $4.1 \pm 0.7_{\text{stat}} \pm 0.3_{\text{syst}}$ & $(3.1 \pm 1.0_{\text{stat}} \pm 0.7_{\text{syst}}) \times 10^{3}$ & - & [9] \\
        12 & & MAGIC, 2017$^{\dagger}$ & flare & PLEC & $2.11\pm 0.14$ & $(16.1\pm 2.3)\times 10^{3}$ & $0.56\pm 0.11$ & [10] \\
        13 & & MAGIC/Fermi-LAT, 2017$^{\dagger \dagger}$ & flare & PLEC & $2.05\pm 0.03$ & $(41.7\pm 2.2)\times 10^{3}$ & $0.49\pm 0.03$ & [10] \\
        \hline
        14 & \multirow{3}{*}{IC~310} & MAGIC, 2009-2010 & quiescent & SPL &  $1.95 \pm 0.12_{\text{stat}} \pm 0.20_{\text{syst}}$ & $6.08 \pm 0.37_{\text{stat}} \pm 1.10_{\text{syst}}$ & -  & [11] \\
        15 &  & MAGIC, 2012-2013 & quiescent & SPL &  $2.36 \pm 0.30_{\text{stat}} \pm 0.15_{\text{syst}}$ & $3.12 \pm 0.91_{\text{stat}} \pm 0.34_{\text{syst}}$ & -  & [12] \\
        16 &  & MAGIC, 2012 & flare & SPL &  $1.51 \pm 0.06_{\text{stat}} \pm 0.15_{\text{syst}}$ & $442.0 \pm 21.0_{\text{stat}} \pm 49.0_{\text{syst}}$ & -  & [12] \\
        \hline
        17 & \multirow{1}{*}{3C~264} & VERITAS, 2018 & quiescent & SPL &  $2.20 \pm 0.27_{\text{stat}} \pm 0.20_{\text{syst}}$ & $1.94 \pm 0.35_{\text{stat}} \pm 0.58_{\text{syst}}$ & - & [13] \\
         \hline

    \end{tabular}
    }
    \begin{tablenotes}
        \item Notes: $^*$Pivot energy $E_0$ at 100 GeV, $^{\dagger}$$E_0 =$ 300 GeV, $^{\dagger \dagger}$$E_0 =$ 198.21 GeV. \\ \textsc{References}.---    [1] \citep{2006Sci...314.1424A}, [2] \citep{2012A&A...544A..96A}, 
        [3] \citep{2008ApJ...679..397A}, [4] \citep{2012AIPC.1505..586B}, [5] \citep{bangale2019study}, 
        [6] \citep{2008ApJ...685L..23A}, [7] \citep{2010ApJ...716..819A}, [8] \citep{2012ApJ...746..141A}, 
        [9] \citep{2012A&A...539L...2A}, [10] \citep{2018A&A...617A..91M}, 
        [11] \citep{aleksic2014rapid},    [12] \citep{ahnen2017first}, [13] \citep{2020ApJ...896...41A}.
        \end{tablenotes} 
    \label{tab:RadGals}
\end{table}

\section{HAWC Data}

The High Altitude Water Cherenkov (HAWC) observatory is a gamma-ray WFD located near Sierra Negra in Puebla, Mexico at an altitude of 4100 m. It is dedicated to the study of VHE events ranging from hundreds of GeV up to hundreds of TeV. HAWC is an extensive air-shower array consisting of 300 water Cherenkov detectors (WCDs) each of which contains a plastic bladder filled with $\sim200,000$ liters of purified water and four photo-multiplier tubes anchored at the bottom, covering a geometrical area of approximately $220,000\;\rm{m}^{2}$ \citep{2023NIMPA105268253A}. With a duty cycle exceeding 95\% and an instantaneous FoV of 2 sr, HAWC covers approximately $2/3$ of the sky \citep{2017ApJ...843...40A}.

Currently, HAWC's reconstruction algorithms are in their PASS 5 version, providing an improvement for the background rejection and angular resolution for gamma rays with energies from $\sim300\,\rm{GeV}$ up to hundreds of TeV as shown in the most recent measurements of the Crab Nebula \citep{2024ApJ...972..144A}. These improvements are relevant in the study of AGNs since their spectra are heavily attenuated in the TeV energy range due to interactions with the Extragalactic Background Light (EBL), meaning that for the farthest AGNs most of their emission is expected at lower energies.

We analyzed 2141 days of PASS 5 HAWC data to perform a search for gamma-ray emission within the regions around the radio galaxies M87, NGC~1275, IC~310 and 3C~264, the search is performed around the positions reported for each radio galaxy in TeVCAT\footnote{Online catalog for TeV sources: \texttt{\hyperlink{http://tevcat.uchicago.edu/}{http://tevcat.uchicago.edu/}}} \citep{2006Sci...314.1424A,2019ApJS..242....5X,2010ATel.2510....1M,2017ApJS..233....3T}. For the spectral analysis we used the Neural Network energy estimator \citep{hawcebinpaper,2024ApJ...972..144A}. Furthermore, we present the results of the long-term monitoring of the aforementioned radio galaxies to explore their behavior at different time scales, with an extended data set covering $\sim 7.5$ years of data (January 2015-June 2022).

\section{Analysis Results}

Based on the results of past observations of the radio galaxies by other gamma-ray observatories presented in Table~\ref{tab:RadGals}, we use the spectral hypothesis reported for each source to produce the significance maps in order to find the maximum significance: a SPL (Eq. \ref{eq:1}) for M87, IC~310 and 3C~264, and a PLEC (Eq \ref{eq:2}) for NGC~1275.

\begin{equation}
    \frac{dN}{dE} = N_{0}\left(\frac{E}{E_{0}}\right)^{-\alpha}\rm{,}\label{eq:1}
\end{equation}

\begin{equation}
    \frac{dN}{dE} = N_{0}\left(\frac{E}{E_{0}}\right)^{-\alpha}\exp{\left(-\frac{E}{E_{\rm{c}}}\right)}\rm{.}\label{eq:2}
\end{equation}

\begin{figure*}[ht!]
    \centering
    \includegraphics[width=0.9\textwidth]{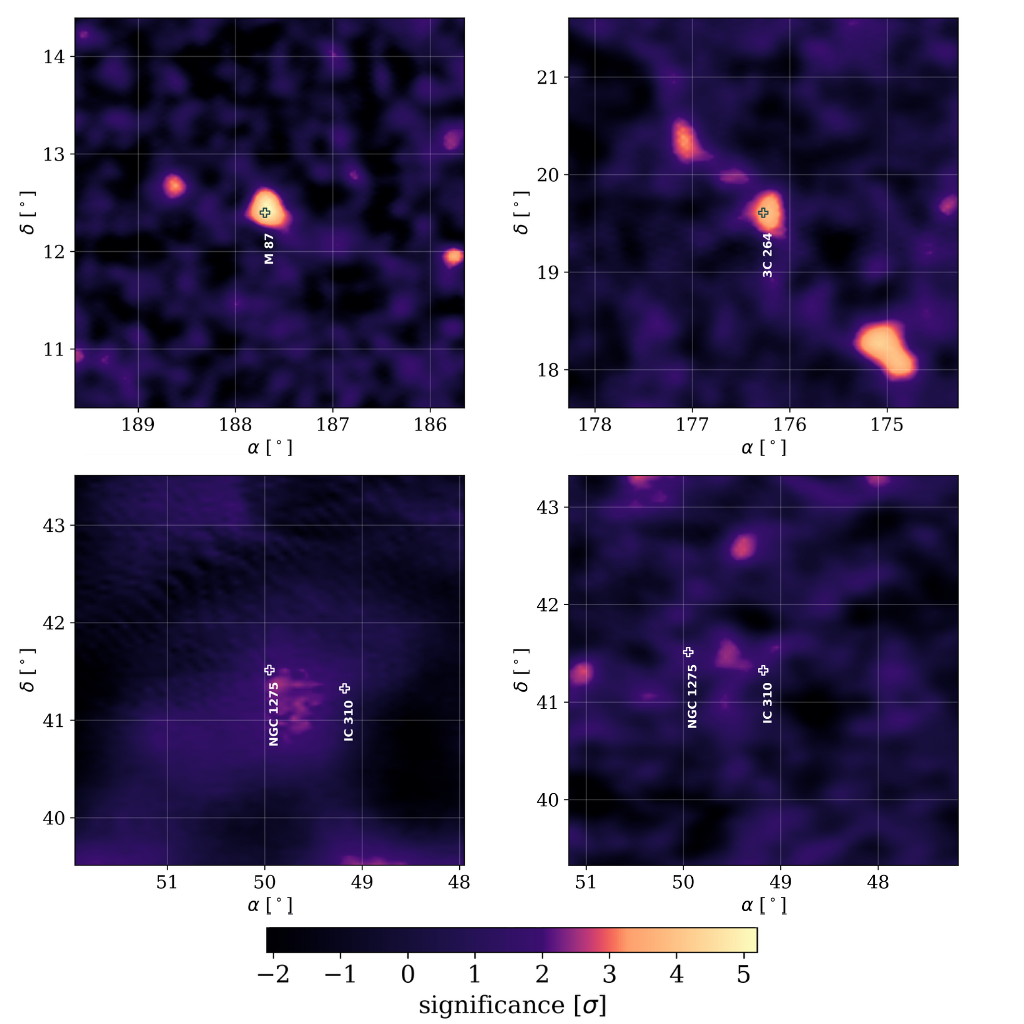}
    \caption{Significance maps obtained with the Neural Network energy estimator. Top left: M87 is detected with a significance of $5.21\sigma$. Top right: 3C~264 is marginally detected with a significance of 3.99 $\sigma$. Bottom left: NGC~1275 is not detected, with a maximum significance of 2.21$\sigma$ at the source position. Bottom right: IC~310 is not detected, with a significance of 1.34$\sigma$ at the source position. Difference between bottom maps is the assumed position for the gamma-ray emission search: around NGC~1275's position in bottom left, and around IC~310's position in bottom right. The elongated region between NGC~1275 and IC~310, visible in both bottom maps, may suggest a contribution from both sources. }
    \label{fig:maps}
\end{figure*}

In Figure \ref{fig:maps}, we show the significance maps and the reported locations of the radio galaxies. M87 stands out as the only source detected with a significance above $5\sigma$ at $(\rm{RA},\rm{Dec})=(187.65^\circ, 12.44^\circ)$, which is consistent with the position of M87 reported in the TeVCAT considering the Point Spread Function (PSF) of HAWC. Regarding the NGC~1275-IC~310 region, none of the two radio galaxies were detected above 3 sigma; with the maximum significance in the NGC~1275 map being 0.19$^\circ$ away from the position reported in TeVCAT for this radio galaxy, while the maximum significance in the IC~310 map being 0.51$^\circ$ away from the position reported in TeVCAT for this radio galaxy. Significance maps for both sources were generated separately with a point source hypothesis, as their angular separation according to TeVCAT is approximately $0.8^\circ$ it is plausible that there is contamination from both sources. 3C~264 is marginally detected with a significance of $3.99\sigma$ at $(\rm{RA},\rm{Dec})=(176.22^\circ,19.59^\circ)$, position consistent with the reported in TeVCAT. 


\begin{figure*}[ht!]
    \centering
    \includegraphics[width=0.77\textwidth]{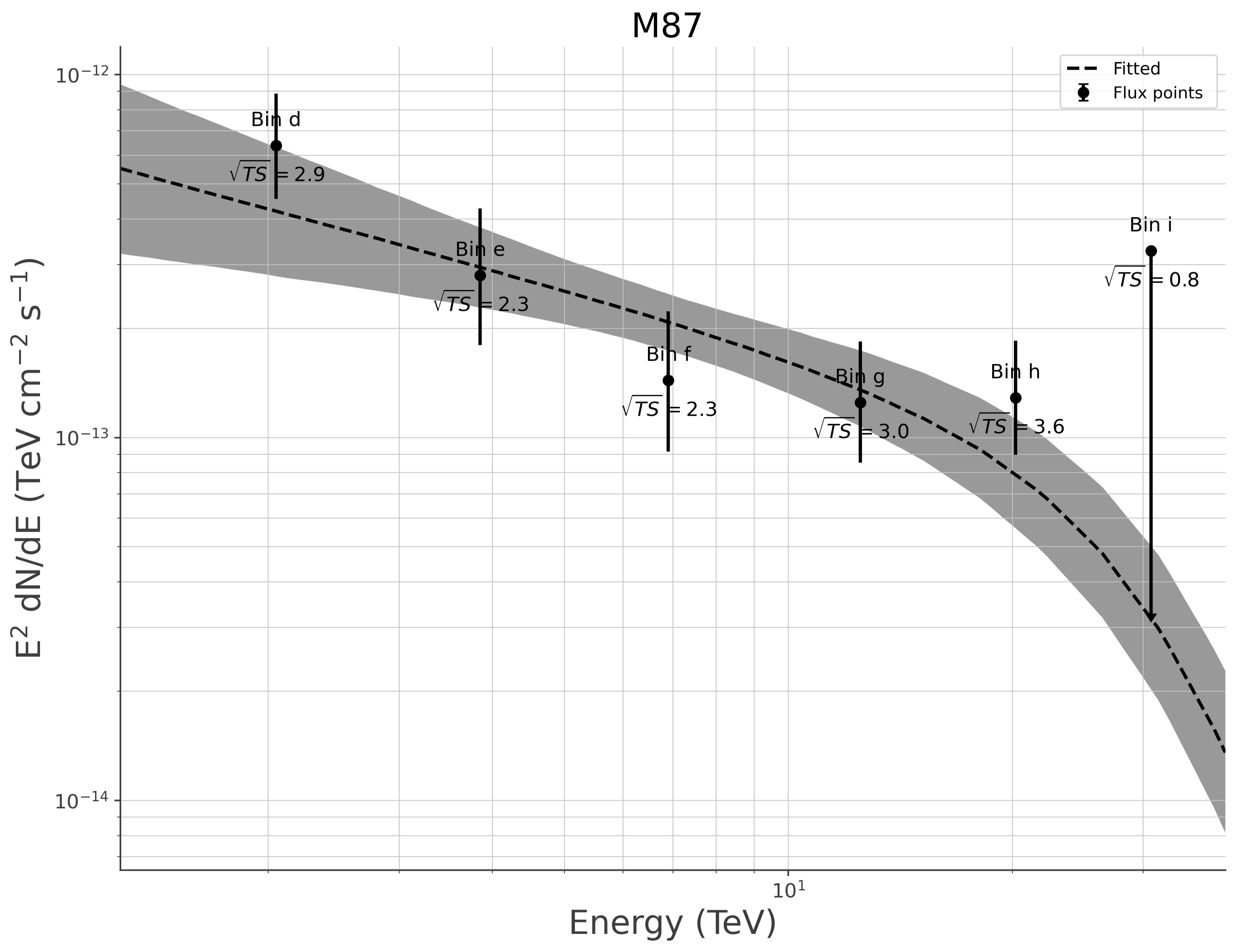}
    \caption{Measured flux (solid black points) and best-fit spectrum (dashed line) of M87. Downward arrows represent upper limits at a $95\%$ confidence level. The fitted spectrum is calculated by assuming a simple power law and fixing the pivot energy and to $E_{p}=12$ TeV. The gray band corresponds to the $1\sigma$ error band of the fit.}
    \label{fig:spectra}
\end{figure*}

Since M87 is the only radio galaxy detected with a significance above $5\sigma$, we analyze its spectrum in detail. The data is best fit to a SPL, where the value for the pivot energy that minimizes correlation between parameters was found to be $12\,\rm{TeV}$ following the methodology described in \cite{Johncrabpaper}. The attenuation caused by the EBL is implemented using the \citet{2008A&A...487..837F} model. Figure \ref{fig:spectra} shows the data and the best fit spectrum for M87, the values obtained for the spectral index and the normalization at $1\,\rm{TeV}$ are $\alpha = 2.53\pm0.29$ and $(7.09\pm 1.24)\times10^{-13}$ $\rm{cm}^{-2}\,{s}^{-1}\,{TeV}^{-1}$ respectively. The maximum energy of the emission observed from M87 by HAWC was calculated using the method implemented in \cite{2017Sci...358..911A}, resulting in 26.5 TeV at a $1\sigma$ level and $15.6$ TeV at $3\sigma$ level. This mean that the emission observed from M87 by HAWC is in the $16-26$ TeV energy range.


\begin{figure}[ht!]
	    \centering
		\includegraphics[width=0.95\textwidth]{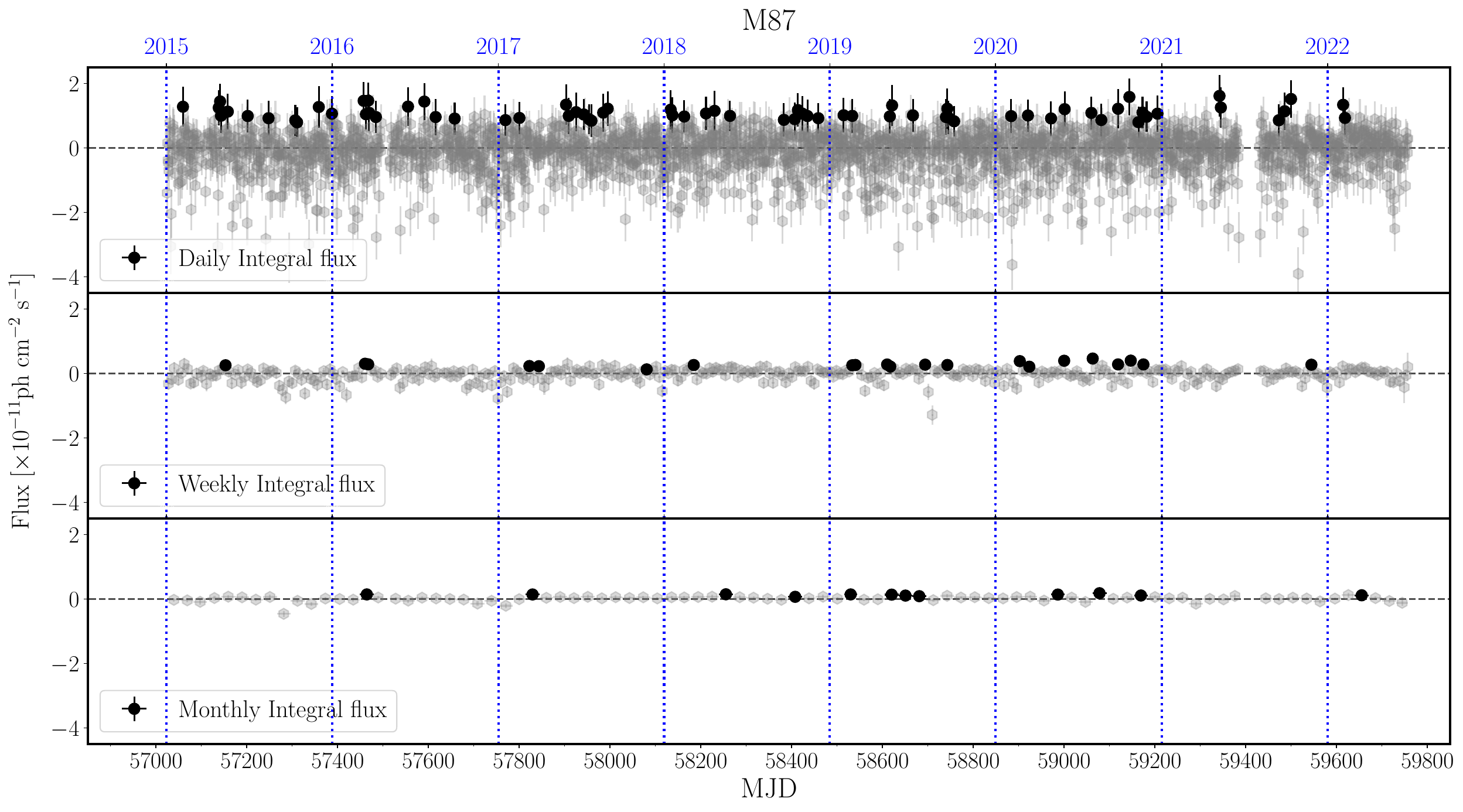}
		\caption{M87 light curves for different time scales. The black point represent the calculated fluxes with a significance of $\geq2\sigma$; while the gray points represent the calculated fluxes with a significance of $<2\sigma$. Top: daily LC. Middle: weekly LC. Bottom: monthly LC.}
		\label{fig:m87hawclc}
	\end{figure}

\begin{figure}[ht!]
	    \centering
		\includegraphics[width=0.95\textwidth]{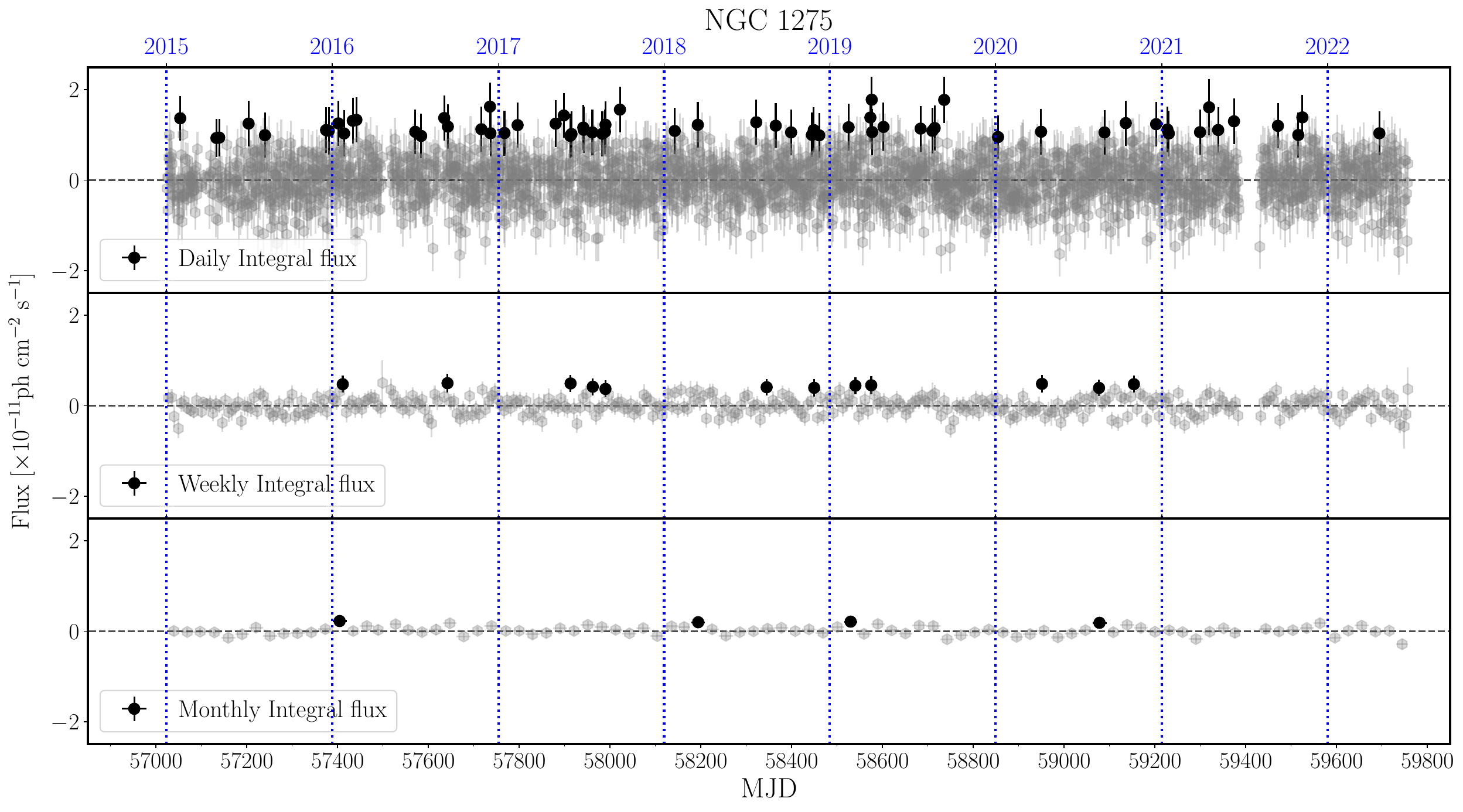}
		\caption{NGC~1275 light curves for different time scales. The black point represent the calculated fluxes with a significance of $\geq2\sigma$; while the gray points represent the calculated fluxes with a significance of $<2\sigma$. Top: daily LC. Middle: weekly LC. Bottom: monthly LC.}
		\label{fig:ngchawclc}
	\end{figure}

\begin{figure}[ht!]
	    \centering
		\includegraphics[width=0.95\textwidth]{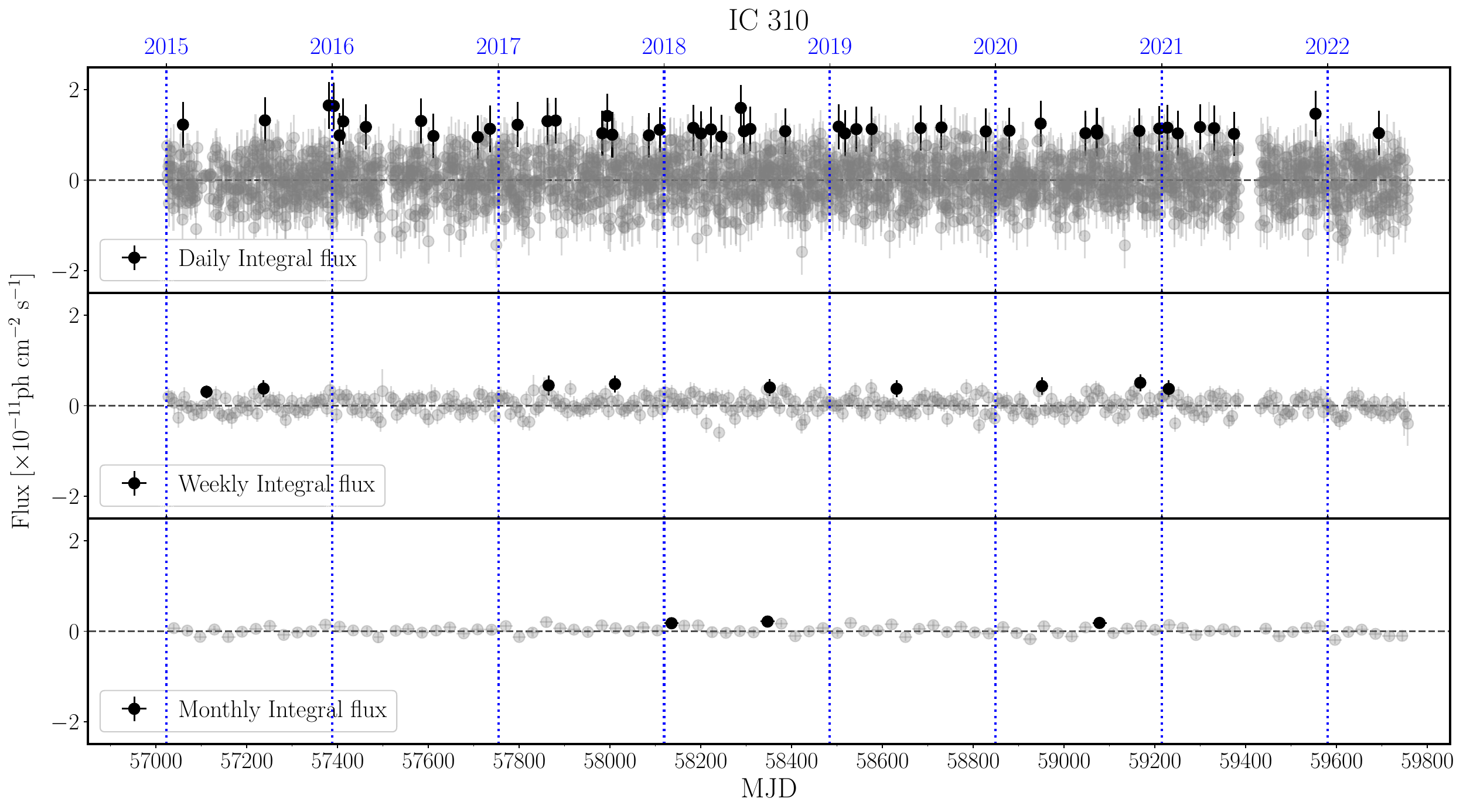}
		\caption{IC~310 light curves for different time scales. The black point represent the calculated fluxes with a significance of $\geq2\sigma$; while the gray points represent the calculated fluxes with a significance of $<2\sigma$. Top: daily LC. Middle: weekly LC. Bottom: monthly LC.}
		\label{fig:ichawclc}
	\end{figure}

\begin{figure}[ht!]
	    \centering
		\includegraphics[width=0.95\textwidth]{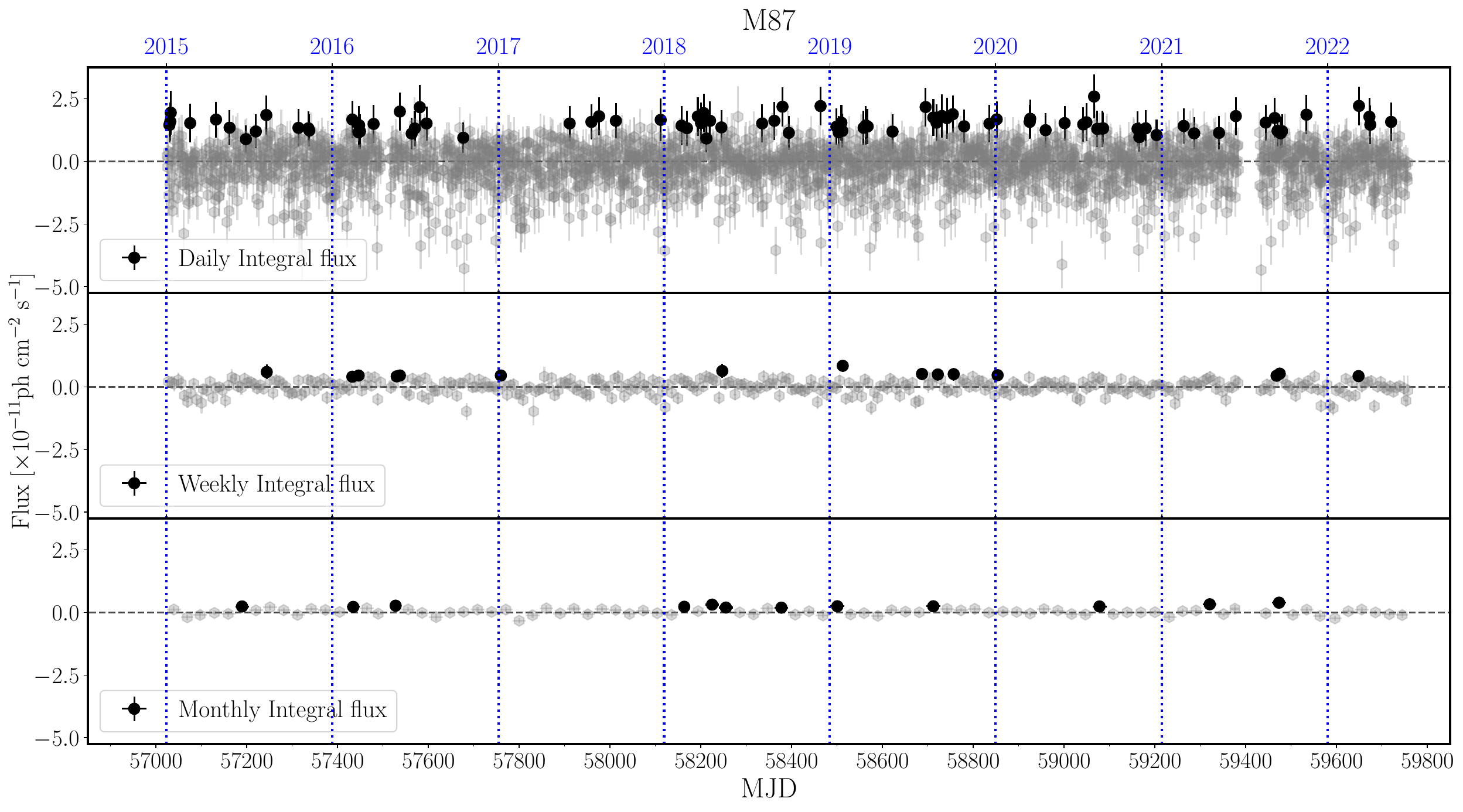}
		\caption{3C 264 light curves for different time scales. The black point represent the calculated fluxes with a significance of $\geq2\sigma$; while the gray points represent the calculated fluxes with a significance of $<2\sigma$. Top: daily LC. Middle: weekly LC. Bottom: monthly LC.}
		\label{fig:3chawclc}
	\end{figure}

We also calculate daily, weekly, and monthly fluxes for each of the four radio galaxies. The light curves are calculated following the same methodology described in \cite{LCRLHAWCPaper}. The spectral hypothesis for the light curves fluxes calculation are the same as the ones employed for the significance maps. Figures \ref{fig:m87hawclc}, \ref{fig:ngchawclc}, \ref{fig:ichawclc} and \ref{fig:3chawclc} show the light curves obtained for M87, NGC~1275, IC~310 and 3C~264 respectively. Only the fluxes with a significance greater than 2$\sigma$ are shown as black data points.

\begin{figure*}[ht!]
	    \centering
		\includegraphics[width=0.75\textwidth]{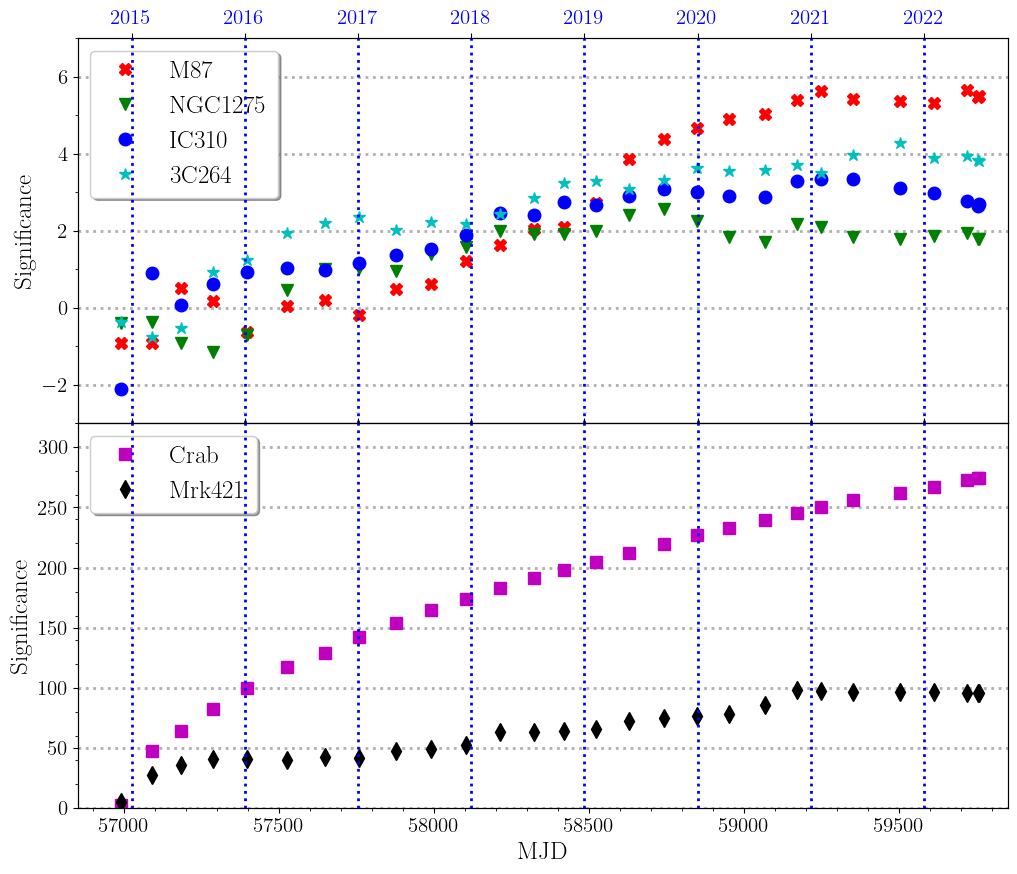}
		\caption{Cumulative significance for the radio galaxies M87 (red crosses), NGC~1275 (blue circles) and 3C~264 (green triangles). For comparison, we also show the cumulative significance for the Crab Nebula (light-blue stars) and Markarian 421 (magenta squares). The behavior of the radio galaxies resembles that of a variable source as Markarian 421 rather than a steady source such as the Crab Nebula.}
		\label{fig:cumulative}
	\end{figure*}

Finally, in order to understand the activity of the sources, we analyze how their significance evolves with respect to observation time. For this, we calculate the significance of the four radio galaxies for different observation time windows, with each time window increasing the observation time by $\sim120$ days, spanning from January 2015 to June 2022. The significance obtained for each increasing time window is refer to as cumulative significance. Results for the cumulative significance for the radio galaxies are shown in the upper plot of Figure \ref{fig:cumulative}.

\section{Discussion}
 
The gamma-ray emission from four radio galaxies have been explored with the HAWC observatory. For M87, the significance map is consistent with a point source and the maximum significance of detection is $5.21\sigma$. As to NGC~1275 and IC~310, the fit with a single point source does not yield a significant detection for any of them. However, the significance maps (Figure \ref{fig:maps}, bottom) show an elongated emission between both sources that does not relates with source confusion due to HAWC's PSF, and seems to extend between both sources. IC~310 has been detected in multiple activity states \citep{2017A&A...603A..25A,graham2019fermi,2024ATel16513....1X} with notably hard spectrum (spectral index $\sim-2$), independently of the level of activity (see Table \ref{tab:RadGals}). Considering that NGC~1275 has a spectrum with a known energy cutoff of around 400-500 GeV, it is more likely that, if TeV emission is observed in this region, it would probably be originated from IC~310 rather than from NGC~1275. Moreover, another radio galaxy, CR~15, is located between NGC~1275 and IC~310 at $(\rm{RA},\rm{Dec})=(49.46^\circ,41.45^\circ)$ \cite{1979A&A....76..109G}. CR~15 has not been reported as a gamma-ray emitter, although it is a strong candidate to contribute to the gamma-ray emission in the region \citep{2023arXiv230903712C}, thus, a contribution from CR~15 cannot be ruled out. Both, IC~310 and CR~15, have been reported as head-tail radio galaxies \citep{sijbring1998multifrequency, ahnen2017first}. Whether the emission shape results from the contribution of one, two, or three of the radio galaxies in the region requires further study as well as more data. With respect to 3C~264, even though the detection is below $5\sigma$, the significance is approximately $4\sigma$ and the hotspot position is consistent with the reported location in the TeVCAT, making this a marginal detection. Moreover, the spectral hypothesis giving the maximum significance for 3C~264 is consistent with the one reported by VERITAS \cite{2020ApJ...896...41A}. 

The energy spectrum of M87 is well fitted by a SPL and agrees with the low-activity spectra reported by different past IACTs' observations \citep{2008ApJ...679..397A, 2012A&A...544A..96A}, indicating that the emission from M87 measured by HAWC is consistent with a low activity state over the whole 2141 days of observation. It is worth to mention that this is the first time that emission above 10 TeV have been observed in a low-activity state in M87, since it was only observed at such energies during the flare episodes of 2005 by H.E.S.S. \citep{2006Sci...314.1424A}. The main difference between HAWC and H.E.S.S. observations is the spectral index, being harder ($\alpha = 2.22\pm 0.15$) for the H.E.S.S. data in comparison with the one fitted for the HAWC data in this analysis ($\alpha = 2.53\pm0.29$) which is consistent with previous reported values for a quiescent state. Furthermore, HAWC's measured spectrum, as well as the maximum energy of the photons, are consistent with the ones reported by LHAASO \citep{2024ApJ...975L..44C}.

The activity of the four radio galaxies has been monitored through the calculation of light curves at daily, weekly and monthly time scales. M87 seems to have slightly more days with significant activity, and, when looking at the weekly and monthly light curves, this activity appears to be more frequent in 2019 and 2021. This is not apparent for NGC~1275 and IC~310, however, there might be a similar trend for 3C~264, particularly in 2018 and 2019. Thus, we interpret this as steady low activity on a weekly time scale in M87, starting in 2019, as shown by the number and distribution of weekly and monthly fluxes. However, it is surprising that, in the case of daily fluxes, 3C~264 exhibits a similar behavior to M87 with a comparable excess in the number of days with significance greater than 2$\sigma$, despite being marginally detected on the significance map of Figure \ref{fig:maps}. This excess ($>70$ days) is higher than the expected background fluctuations at the same significance level ($\sim59$ days) for the observation time. It is worth noticing that for M87, with a spectral normalization of the order of $10^{-13}\;\text{cm}^{-2}\;\text{s}^{-1}$ at $1\;\rm{TeV}$  and a spectral index of $\alpha = 2.53\pm0.29$, the resulting integral flux above 1 TeV in the daily light curves is of the order of $10^{-11}\;\text{cm}^{-2}\;\text{s}^{-1}$. This is up to two orders of magnitude higher than the one obtained by integrating the spectrum obtained with the overall HAWC data of Figure \ref{fig:spectra}, or even the reported spectral parameters from Table \ref{tab:RadGals}. Therefore, M87 and 3C~264 showed variations in their activity resulting in high flux increments at daily time scales. Nevertheless, HAWC's data do not exhibit the 2022 flare reported by LHAASO for M87 \citep{2024ApJ...975L..44C}.

The cumulative significance plots show the evolution of the significance over observation time with HAWC. The final value for the cumulative significance is in agreement with the one obtained for each radio galaxy in the significance maps which covers 2141 days of data. For comparison, we also show, in the bottom plot of Figure \ref{fig:cumulative}, the corresponding cumulative significance for the Crab Nebula (a constant steady source) and for Markarian 421 (a variable source) using HAWC data spanning the same time period as for the radio galaxies. The general behavior of the cumulative significance over time, in the four radio galaxies, resembles the behavior of a variable source such as Markarian 421. It is evident that M87 started to increase its emission (and thus, its significance) resulting in a $5\sigma$ detection by 2021. NGC~1275 and 3C~264 seem to have increased its emission (and its significance) around early 2017 and early 2016 respectively; but their flux was not high enough to be detected by HAWC.

\section{Conclusions}

We have analyzed the TeV emission of four radio galaxies within the FoV of the HAWC observatory using 2141 days of PASS~5 reconstructed data. We detected gamma-ray emission from M87 with a significance of $5.21\sigma$. 3C~264 was marginally detected, with a significance of approximately $4\sigma$. In contrast, NGC~1275 and IC~310 were not detected, both showing a significance below $3\sigma$. Interestingly, the significance maps for NGC~1275 and IC~310 show an extended region containing both radio galaxies. This region might be of interest as it contains three potential TeV emitters: NGC 1275, IC 310, and the radio galaxy CR 15, located between the two. In future works, it will be worth to further analyze this region in detail, taking into account the possible different contributions from these sources.
    
Regarding the energy spectrum of M87, the measured flux suggest that this radio galaxy has been in a low-activity state compared to previous observations by IACTs. Nevertheless, the spectrum extends up to 20 TeV; such energies were only observed during the flares episodes of this radio galaxy in 2005. The spectrum is well-fitted by a simple power-law with spectral index of $\alpha = 2.53\pm0.29$, a softer spectrum in comparison to the ones reported in flare activity. The maximum energy estimated for the emission observed by HAWC at different significances indicates that the observed emission is in the $16-26$ TeV energy range.

Daily, weekly, and monthly light curves were obtained for M87, NGC~1275, 3C~264 and IC~310 using PASS~5 data spanning over $\sim7.5$ years. This is the first, uninterrupted very long term TeV monitoring of radio galaxies that is not biased to a particular activity state. There appears to be slightly more significant emission in M87, when observing at weekly scales, starting at 2019. There might be a similar trend in 3C~264, seemingly starting around early 2018 at weekly and monthly scales. While the daily light curve for 3C~264 shows a number of flux points comparable with those in the M87 light curve, the number of weeks with emission above $2\sigma$ is significantly higher for M87 than for 3C~264. This can be interpreted as 3C~264 having a high steady emission for shorter time periods than M87, resulting in an overall lower flux and, thus, a non-positive detection of 3C~264 over the full observation time with HAWC. It should be noted that M87 and 3C~264 are extremely variable sources as previous observations had shown\citep{2006Sci...314.1424A,2020ApJ...896...41A}, where the detection of 3C~264 by VERITAS was associated with an enhancement of its flux due to its VHE variability.

Finally, the cumulative significance clearly shows that the radio galaxies studied in this work exhibit behavior similar to that expected for a variable gamma-ray source such as the blazar Mrk 421. This might be in accordance with some unification models where radio galaxies are just BL Lac objects with their jet misaligned to our line-of-sight \citep{1995PASP..107..803U,2012MNRAS.420.2899G,2024Ap&SS.369..106I}, expecting a similar behavior from both sources. Thus, monitoring radio galaxies to identify increments on their activity, as well as flare episodes, may give further insights into the emission processes of AGNs. Furthermore, blazars are expected to contribute to the diffuse neutrino emission \citep{2020PhRvL.125l1104A}, supporting importance of monitoring radio galaxies as identifying periods of higher activity and flares can be used as a complement for multi-messenger studies. The HAWC observatory continuously collects data and will continue monitoring these radio galaxies for further studies.

\begin{acknowledgments}
We acknowledge the support from: the US National Science Foundation (NSF); the US Department of Energy Office of High-Energy Physics; the Laboratory Directed Research and Development (LDRD) program of Los Alamos National Laboratory; Consejo Nacional de Ciencia y Tecnolog\'{i}a (CONACyT), M\'{e}xico, grants LNC-2023-117, 271051, 232656, 260378, 179588, 254964, 258865, 243290, 132197, A1-S-46288, A1-S-22784, CF-2023-I-645, CBF2023-2024-1630, c\'{a}tedras 873, 1563, 341, 323, Red HAWC, M\'{e}xico; DGAPA-UNAM grants IG101323, IN111716-3, IN111419, IA102019, IN106521, IN114924, IN110521 , IN102223; VIEP-BUAP; PIFI 2012, 2013, PROFOCIE 2014, 2015; the University of Wisconsin Alumni Research Foundation; the Institute of Geophysics, Planetary Physics, and Signatures at Los Alamos National Laboratory; Polish Science Centre grant, 2024/53/B/ST9/02671; Coordinaci\'{o}n de la Investigaci\'{o}n Cient\'{i}fica de la Universidad Michoacana; Royal Society - Newton Advanced Fellowship 180385; Gobierno de España and European Union-NextGenerationEU, grant CNS2023- 144099; The Program Management Unit for Human Resources \& Institutional Development, Research and Innovation, NXPO (grant number B16F630069); Coordinaci\'{o}n General Acad\'{e}mica e Innovaci\'{o}n (CGAI-UdeG), PRODEP-SEP UDG-CA-499; Institute of Cosmic Ray Research (ICRR), University of Tokyo. H.F. acknowledges support by NASA under award number 80GSFC21M0002. C.R. acknowledges support from National Research Foundation of Korea (RS-2023-00280210). We also acknowledge the significant contributions over many years of Stefan Westerhoff, Gaurang Yodh and Arnulfo Zepeda Dom\'inguez, all deceased members of the HAWC collaboration. Thanks to Scott Delay, Luciano D\'{i}az and Eduardo Murrieta for technical support.
\end{acknowledgments}

\bibliography{biblio}{}
\bibliographystyle{aasjournal}



\end{document}